\documentstyle{elsart}
\begin{document}
\begin{frontmatter}

%\bibliographystyle{unsrt}

%\draft
%\widetext
\title{Experimental evidence of a metal-insulator transition
in~a~half-filled~Landau~level
}
\author{C.--T.~Liang, J.~E.~F.~Frost, M.~Y.~Simmons,
D.~A.~Ritchie, and M.~Pepper
}
\address{
Cavendish Laboratory,
Madingley Road,
Cambridge CB3 0HE,
United Kingdom}

\date{\today}

\maketitle

%\widetext
\begin{abstract} 
%\leftskip 54.8pt
%\rightskip 54.8pt

We have measured the low-temperature transport properties of 
a high-mobility front-gated 
GaAs/Al$_{0.33}$Ga$_{0.67}$As heterostructure. By changing the
applied gate voltage, we can vary the amount of disorder within the
system. At a Landau level filling factor $\nu =1/2$, where the system can
be described by the composite fermion picture, we observe a
crossover from metallic to insulating behaviour as the disorder is
increased. Experimental results and
theoretical prediction are compared.

%\pacs{PACS numbers: 73.40.Gk, 73.20.Dx, 73.40.-c}
\end{abstract}
\end{frontmatter}
 
\newpage

%\section{Introduction}
The fractional quantum Hall effect (FQHE) \cite{Tsui}
arises from strong electron-electron
interactions, causing the two-dimensional (2D) electrons to condense
into a fractional quantum Hall liquid \cite{Laughlin}.
In the elegant composite fermion (CF) picture, where each electron
is bound to an even number of magnetic flux quanta \cite{Jain}, the FQHE 
can be understood as a manifestation of the integer quantum Hall
effect of weakly interacting composite fermions.  
It has been shown \cite{KZ,HLR} that at Landau
level filling factor $\nu = 1/2$, a 2D electron gas (2DEG) 
can be mathematically transformed into a CF system
interacting with a Chern--Simons gauge field. A wide variety of
recent experimental results
\cite{Du1,Kang,Leadley,Kuku,Liang}
have demonstrated that at $\nu = 1/2$ the effective magnetic field
acting on the composite fermions is zero. 

It is well known that both weak localisation and electron-electron
interaction theories \cite{WLEEI} produce a logarithmic dependence of
conductivity with temperature observed \cite{Uren} in weakly disordered
2D electron systems at zero magnetic field. 
Rokhinson, Su, and Goldman \cite{Rokhinson} 
ascribe the logarithmic temperature $T$ dependence
of conductivity at $\nu = 1/2$
to CF--CF interactions, analogous to electron-electron
interactions at zero magnetic field. Recent theoretical work \cite{Khvesh}
has shown that in the case of short--range interactions, the correction
term to the classical composite fermion conductivity due to CF--CF
interactions is given by

\begin{equation} 
\delta\sigma^{CF}_{xx}=\left(\frac{e^2}{2{\pi}h}\ln(k_{F}\ell)\right)
\ln\left(T\tau_{tr}\right),
\end{equation}

where $k_{F}$ is the CF wave vector, $\tau_{tr}$ and $\ell$ are the
elastic scattering time and elastic mean free path for composite
fermions, respectively. Note that weak localisation is
suppressed at $\nu = 1/2$ since the Chern--Simons gauge field fluctuations
break time reversal symmetry for impurity scattering \cite{HLR,Rokhinson}. 

Recently there has been much interest in the global phase diagram in 
the quantum Hall effect \cite{KLZ,Jiang2}. With the same
theoretical approach \cite{KLZ},
Kalmeyer and Zhang \cite{KZ} have investigated 2D electron systems at
$\nu = 1/2$ with various amount of disorder. They predict that for
sufficiently weak disorder, the system is metallic at $\nu = 1/2$,
characterised by positive magnetoresistance (PMR) centred around
$\nu = 1/2$. When disorder increases, the system enters an
insulating phase where negative magnetoresistance (NMR) around 
$\nu = 1/2$ is observed. In this communication, we shall show that 
our experimental results on the temperature dependence of CF 
conductivities as a function of disorder are consistent with 
their prediction of a metal-insulator transition.
However, we cannot test the prediction of a
crossover from PMR to NMR due to large background magnetoresistance
around $\nu=1/2$ as the disorder becomes stronger.

We have studied gated Hall bars made from a high-mobility
GaAs/Al$_{0.33}$Ga$_{0.67}$As heterostructure, with the magnetic field $B$
directed perpendicular to the interface. At gate voltage $V_{g}=0$,
the carrier concentration of the 2DEG is 
$\approx 1.0\times 10^{15}$~$\mathrm{m^{-2}}$
with a mobility of $300$~$\mathrm{m^{2}/Vs}$ without illumination.
Experiments were performed in a $^{3}$He cryostat at 0.3 K using standard
four-terminal ac phase-sensitive techniques. By changing $V_{g}$, we
can vary the carrier density and hence effectively the disorder in our
system. Three samples showed similar characteristics, and
measurements taken from one of these are presented in this paper.
 
Figure~1 shows the four-terminal longitudinal $\rho_{xx}$ and
transverse $\rho_{xy}$ magnetoresistance for $V_{g}$=0.
The minima in $\rho_{xx}$ coincide with the quantum Hall plateaux, 
indicating that the carrier density $n_{s}$ in our wafer is uniform. 
There is a good fit 
$n_{s} = (2.136 \times 10^{15}V_{g}+9.124 \times 10^{14})\mathrm{m^{-2}}$
over the measurement range --0.38~V $\leq V_{g} \leq$ 0 V. Thus
$n_{s}(V_{g})$ in our system can be described
by a simple capacitor plate model with a distance 0.322 $\mu$m between
the surface Schottky gate and the underlying 2DEG, in close agreement
with the intended as--grown depth of 0.3 $\mu$m. 

Figure~2 shows the magnetoresistance measurements at various $V_{g}$.
At $V_{g} = 0$, PMR around $\nu = 1/2$ is observed. 
As $V_{g}$ is made more negative and hence the amount of disorder
within the system is increased, $\rho_{xx}$ at $\nu=1/2$
increases, and PMR centred around $\nu = 1/2$
gradually diminishes as indicated by arrows. At $V_{g} \leq$ --0.22 V,
PMR around $\nu = 1/2$ is no longer observable and $\rho_{xx}(B)$
increases with magnetic field, as shown in Fig.~2 and the inset. The linear
rising background $\rho_{xx}$ \cite{Stormer} may mask the
magnetoresistance around $\nu=1/2$, excluding the possibility of
testing the theoretical prediction of a PMR/NMR crossover \cite{KZ}.
For $V_{g} = -0.34$ and $-0.38$ ~V, the magnetoresistance minima at 
$\nu = 1/3$, the most pronounced fractional quantum Hall state, 
are observed at around 2.7~T and 1.6~T, respectively, demonstrating 
that the composite fermion picture is valid in this case.
 For $V_{g} = -0.412$~V and $-0.413$~V, the
minimum at $\nu = 1/3$ is no longer observable. This is consistent
within the picture of the global phase diagram \cite{KZ,KLZ}. As the amount of
disorder in the system is sufficiently strong, the $\nu =1/2$
state enters an insulating phase and the fractional quantum Hall
state $\nu =1/3$ is no longer observable, as shown in the inset to Fig.~3~(a).

Existing measurements  
\cite{Du1,Kang,Leadley,Kuku,Liang}
were mostly performed in the regime where PMR around $\nu = 1/2$ is
clearly observed. We choose to concentrate on the regime where PMR around
$\nu = 1/2$ disappears
and we have also studied the $T$ dependence of CF conductivity 
in this limit. Note that composite fermion conductivity
$\sigma^{CF}_{xx}$ is given by 1/$\rho_{xx}$ \cite{HLR,Du1}. 
Figure 3~(a) shows the 
composite fermion conductivity for $V_{g} =$--0.34~V at various
temperatures. There is a good logarithmic fit
$\sigma^{CF}_{xx}$= ($3.663 \times $ln $T$ + $124.363$) $\mu$S 
for 0.3 K $\leq T \leq$ 0.65 K. As shown
in Fig. 3~(b), there is also a good fit 
$\sigma^{CF}_{xx}$= ($6.909 \times $ln $T$ + $69.457$) $\mu$S
for 0.3 K $\leq T \leq$ 0.55 K at $V_{g} =$--0.38~V, where
the disorder within the system is further increased. We observe that
$\sigma^{CF}_{xx}$ shows a deviation from a logarithmic temperature
dependence at high temperatures, similar to those reported in previous 
work \cite{Kuku,Rokhinson}. This effect might be due to CF-phonon scattering 
\cite{Kang2}. Using equation (1), we estimate the prefactor in ln $T$
term to be 7 for $V_{g} =$--0.38~V, in excellent agreement with the
experimental results. However, for $V_{g} =$--0.34~V, the estimated
prefactor is 11.4, somewhat larger than the value 3.663 obtained from
the experimental data. It is important to note that the theory predicts 
the prefactor in ln $T$ increases as $k_{F}\ell$ increases, i.e., 
when the disorder becomes weaker. However, our experimental results
show the opposite behaviour. The discrepancy is not understood at
present.

Previous experimental results have shown that
$\sigma^{CF}_{xx}(\nu=1/2)$ exhibits a weak \cite{Du1,Leadley} or
logarithmic \cite{Kuku,Rokhinson} temperature dependence, 
demonstrating the existence of a composite fermion metallic phase.
Figure~4 shows $\sigma^{CF}_{xx}$($T$) for $V_{g} =$ --0.412~V and 
--0.413~V. The
experimental data follows 2D Variable Range Hopping (VRH) 
$\sigma^{CF}_{xx} \approx$ $\mathrm{exp}$$ [-(T_{0}/T)^{1/3}]$,
with a fit yielding characteristic temperatures $T_{0}$ of 24~K and 
25~K for $V_{g} = $ --0.412~V and --0.413~V, respectively.
The exponential temperature dependence of $\sigma^{CF}_{xx}$
provides, to the best of our knowledge, the first experimental
evidence of an insulating phase for CFs at $\nu = 1/2$.
The data shown in Fig. ~3 and 4 
suggests that a transition from metallic (ln $T$ dependence) to insulating
($\mathrm{exp}$$[-(1/T)^{1/3}]$ dependence) behaviour
occurs as the amount of disorder is
increased. To map out the proposed phase diagram \cite{KZ,HLR,KLZ},
further detailed measurements on a large number of samples
with varying degree of disorder and different carrier
densities need to be performed.

In conclusion, we have studied a 2D composite fermion system with
various of amount of disorder. At $\nu = 1/2$, a crossover from metallic to
insulating behaviour is observed in the temperature dependence of the
CF conductivity, in agreement with theory. However, the
rapidly rising background magnetoresistance
around $\nu = 1/2$ precludes the observation of 
a crossover from positive to negative
magnetoresistance with increasing disorder.

This work was funded by the United Kingdom
(UK) Engineering and Physical Sciences Research Council. We thank
J.D.F. Franklin, A.R. Hamilton, C.G. Smith and Chang-Ming Ho 
for helpful discussions, K.J. Thomas for
sample preparation, and N.P.R. Hill for experimental assistance.
C.T.L. acknowledges financial support from
Hughes Hall College, the C.R. Barber Trust Fund, and the Committee of
Vice--Chancellors and Principals, UK.

%\begin{references}

\centerline{\bf Figure Captions}

Figure 1.
%\begin{figure}
%\epsfxsize=70truemm
%\centerline{\epsffile{jeff1.ps}}
%\caption {
Longitudinal $\rho_{xx}$ and transverse $\rho_{xy}$ magnetoresistance
measurements for $V_{g}$ = 0 V. 
%} \label{1} \end{figure}

Figure 2.
%\begin{figure}
%\epsfxsize=70truemm
%\centerline{\epsffile{jeff2.ps}}
% \caption { 
Magnetoresistance measurements  $\rho_{xx}(B)$ for various $V_{g}$.
From left to right: $V_{g}$ = --0.3 V to --0.08V in 0.02 V steps,
$V_{g}$ =
--0.04 V, and $V_{g}$ = 0 V, respectively. The inset shows
$\rho_{xx}(B)$ for
$V_{g}$ = --0.34 V, --0.38 V, --0.412 V and --0.413 V, respectively.
The positions of $\nu = 1/2$ are indicated by arrows.    
%} \label{2} \end{figure}

Figure 3.
%\begin{figure} 
%\epsfxsize=80truemm 
%\centerline{\epsffile{jeff3.ps}}
% \caption {
(a) $\sigma^{CF}_{xx}(T)$ at $\nu = 1/2$ for $V_{g} =$--0.34~V.
(b) $\sigma^{CF}_{xx}(T)$ at $\nu = 1/2$ for $V_{g} =$--0.38~V. 
The straight line fits are discussed in the text. 
The inset shows a schematic global phase diagram illustrating that
in a highly disordered system, 
as the magnetic field is increased,
at $\nu=1/2$ the systems can enter an insulating (I) phase rather
than a metallic (M) phase and the fractional quantum Hall state
$\nu=1/3$ may be no longer observable as shown by the dotted line.
 
%} 
%\label{3} \end{figure}

Figure 4.
%\begin{figure} 
%\epsfxsize=80truemm 
%\centerline{\epsffile{jeff4.ps}}
% \caption {
$\sigma^{CF}_{xx}(T)$ at $\nu = 1/2$ for $V_{g} =$--0.412~V (marked by
circles) and for $V_{g} =$--0.413~V (marked by squares).
The linear fits are discussed in the text.
%} 
%\label{4} \end{figure}

%\end{multicols}
\end{document}